\documentclass[twocolumn,tighten,times]{aastex631}

\usepackage{bm}

\begin{document}

\title{Alfv{\'e}n number for the Richtmyer-Meshkov instability in
  magnetized plasmas}

\correspondingauthor{Takayoshi Sano}
\email{sano@ile.osaka-u.ac.jp}

\author[0000-0001-9106-3856]{Takayoshi Sano}
\affiliation{Institute of Laser Engineering, Osaka University, Suita,
  Osaka 565-0871, Japan}

\begin{abstract}

Magnetohydrodynamical evolution of the Richtmyer-Meshkov instability (RMI) is investigated by two-dimensional MHD simulations.
The RMI is suppressed by a strong magnetic field, whereas the RMI amplifies an ambient magnetic field by many orders of magnitude if the seed field is weak. 
We have found that the suppression and amplification processes can be evaluated continuously along with the amplitude of the Alfv{\'e}n number $R_A$, which is defined as the ratio of the linear growth velocity of the RMI to the Alfv{\'e}n speed at the interface.
When the Alfv{\'e}n number is less than unity, the Lorentz force acting on the fluid mitigates the unstable motion of the RMI significantly, and the interface oscillates stably in this limit.
If $R_A \gtrsim 1$, on the other hand, the surface modulation increases due to the growth of the RMI. The maximum strength of the magnetic field is enhanced up to by a factor of $R_A$.
This critical feature is universal and independent of the initial Mach number of the incident shock, the Atwood number, corrugation amplitude, and even the direction of the initial magnetic field. 

\end{abstract}



\section{Introduction}

The Richtmyer-Meshkov instability (RMI) is one of the interfacial instabilities induced by shock interaction \citep{richtmyer60,meshkov69}.
When a shock front passes through a spatially corrugated contact discontinuity, the interface is generally subjected to the RMI \citep{brouillette02,nishihara10,zhou17}.
The RMI plays an essential role in the propagation of supernova shocks through the inhomogeneous interstellar matters, which could excite interstellar turbulence and enhance ambient magnetic fields \citep[e.g.,][]{giacalone07,inoue09}.
Thus the understanding of magnetohydrodynamical characteristics of the RMI has a quite important meaning to reveal star formation events in the turbulent interstellar medium \citep{mckee07}.

At the linear phase of the RMI, the shock acceleration of the interface is impulsive and causes the perturbation amplitude to grow linearly in time.
\citet{richtmyer60} has proposed a simple estimate of the growth rate of the amplitude, which is obtained by a generalization of the Rayleigh-Taylor formula with an impulsive acceleration.
Then, \citet{meyer72} observed that the Ricthmyer prescription should be modified using an averaged value between the pre-shocked and post-shocked interface amplitude in order to obtain agreement between the numerical solution and the linear theory. 
Unfortunately, these empirical prescriptions are likely to fail for high compressions \citep{vandenboomgaerde98}.
The more accurate formula for the asymptotic growth velocity is derived by \citet{wouchuk96,wouchuk97}, in which the tangential velocities generated at the interface are the key to determine the evolution of the RMI.   
The origin of the parallel velocity to the interface is the refracted flow caused by the oblique angle of the transmitted shock and the reflected shock or rarefaction. 

In this paper, we investigate the suppression of the RMI due to the presence of a strong magnetic field.
The stabilization of the RMI could be realized by several situations such as a gradual density transition at the interface \citep{sano20} and freeze-out \citep{wouchuk15}.
The Lorentz force working on the interface can stabilize the unstable growth of the RMI \citep{samtaney03}.
The motivation here is to empirically derive a critical condition for the magnetic field strength, which may be helpful to predict the nature of the interstellar turbulence and design plasma experiments for the demonstration \citep{sano21}.

Analytical modeling of the MHD RMI is cumbersome because the unperturbed state is time-dependent. 
Thus, numerical simulations have been a suitable tool to examine the role of magnetic fields on the RMI, even in the linear phase.
Depending on the field direction relative to the incident shock surface, the magnetic field works differently on the fluid motions of the RMI. 
When the magnetic field is aligned with the shock propagation, the suppression of the instability is caused by changes in the shock refraction process at the contact discontinuity.
The role of the magnetic field is to prevent the deposition of circulation on the interface \citep{wheatley05,wheatley09}. 
When the magnetic field is parallel to the shock surface, the instability can be mitigated by the Lorentz force \citep{cao08}.
The magnetic field component perpendicular to both the shock propagation and interface has little influence on the linear growth of the RMI \citep{shen20}.

In the previous works \citep{sano12,sano13}, the cases when an incident shock travels from lighter fluid to heavier one (shock-reflected cases) are intensely discussed.
However, the growth of the RMI takes place in the opposite situation, that is, rarefaction-reflected cases. 
Then, in this work, we perform a systematic survey of the rarefaction-reflected RMI, including an ambient uniform magnetic field.
Thanks to the comprehensive survey, it turns out that a robust critical condition is described by the Alfv{\'e}n number, which is given by a ratio of the linear growth velocity of the RMI and the Alfv{\'e}n speed.

The plan of this paper is as follows.  
In \S 2, the numerical setup for the single-mode analysis of the RMI is described.
Theoretical models for the linear growth velocity of the RMI are briefly reviewed in \S 3.
We also demonstrate the importance of heavy-to-light configuration for the RMI. 
In \S 4, numerical results of MHD evolution of the RMI are explained based on the amplitude of the Alfv{\'e}n number. 
Comparisons with the other interfacial instabilities in terms of the role of magnetic fields are discussed in \S 5.
We also discuss more realistic situations introducing a multi-mode surface modulation.
Finally, our conclusion is summarized in \S 6.

\section{Numerical Method}

\subsection{Numerical Setup}

\begin{figure*}
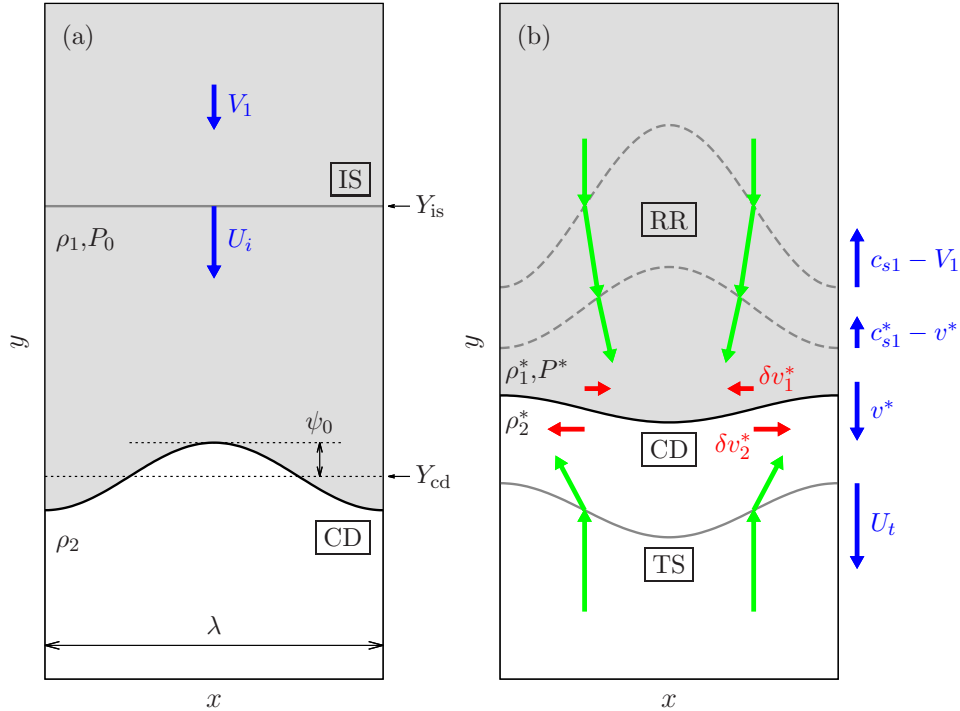

\begin{center}
\includegraphics[scale=0.9,clip]{fig1a.eps}%
\hspace{1mm}
{\includegraphics[scale=0.9,clip]{fig1b.eps}}%
\caption{
(a) 
Schematic picture of the initial configuration for single-mode analysis of the RMI.   
Two fluids are divided by a contact discontinuity (CD). 
The densities of the heavier fluid ``1'' and lighter fluid ``2'' are $\rho_1$ and $\rho_2$, and the uniform pressure for both fluids is $P_0$.
The interface is corrugated sinusoidally with the wavelength $\lambda$ and the amplitude $\psi_0$.  
An incident shock (IS) propagates in the heavier fluid ``1'' with the shock velocity $U_i$.   
Here $V_1$ is the flow velocity behind the incident shock.
(b) 
Sketch of the shock-front shapes after the incident shock hits the corrugated interface. 
The transmitted shock (TS) and reflected rarefaction (RR) travel from the contact discontinuity in the opposite direction.
The pressure and velocity at the contact discontinuity are $P^{\ast}$ and $v^{\ast}$, and the densities behind the transmitted and reflected waves are $\rho^{\ast}_1$ and  $\rho^{\ast}_2$.
The transmitted shock velocity is $U_t$ and the rarefaction wave travels with the local sound speed, for instance, $c_{s1}^{\ast} = (\gamma P^{\ast} / \rho_{1}^{\ast})^{1/2}$. 
Because of the obliqueness of the wavefronts, tangential flows, $\delta v_1^{\ast}$ and $\delta v_2^{\ast}$, are generated at both sides of the interface.
The green arrows at the wavefronts show the refraction features of the fluid motions at the transmitted and reflected waves.
\label{fig1}}
\end{center}
\end{figure*}

Figure~\ref{fig1}(a) illustrates the initial configuration for the single-mode analysis of the RMI when a rarefaction wave is reflected.
Two fluids with different densities, $\rho_1$ and $\rho_2 (< \rho_1)$, are separated by a contact discontinuity at $y = Y_{\rm cd}$. 
A planar shock propagating through the heavier fluid ``1'' ($y > Y_{\rm cd}$) hits the interface at a time $t = 0$. 
Here the $x$- and $y$-axis are perpendicular and parallel to the shock surface.  
The incident shock velocity is $U_i (< 0)$ and the fluid velocity behind the shock is $V_1 (< 0)$.
The pre-shocked pressure $P_0$ is uniform in both fluids.
The Mach number of the incident shock is defined as $M = |U_i|/c_{s1}$ where $c_{s1} = (\gamma P_0/ \rho_1)^{1/2}$ is the sound speed of the fluid ``1''. 
The interface has an initial corrugation of a sinusoidal form, $y = Y_{\rm cd} + \psi_0 \cos (k x)$, where $\psi_0$ is a corrugation amplitude, $k = 2 \pi / \lambda$ is the perturbation wavenumber, and $\lambda$ is the wavelength.   

The initial geometry of a magnetic field is assumed to be uniform with the size of $|\bm{B}| = B_0$ in the pre-shocked region ($y < Y_{\rm is}$). 
As for the field direction, we consider the cases of a parallel MHD shock ($B_y = B_0$) and a perpendicular shock ($B_x = B_0$).
A uniform magnetic field parallel to the incident shock velocity is assumed in the entire region for the parallel shock case.
While, for the perpendicular shock case, the initial field is calculated from the following vector potential to avoid the presence of the normal field at the interface,
\begin{equation}
A_{z} (x, y) = B_{0} y
- B_{0} \psi_0 \cos k x \exp [ - k | y - Y_{\rm cd} (x) | ] \;, 
\end{equation}
using the relation $\bm{B} = \bm{\nabla} \times \bm{A}$.
The physical quantities in the post-shocked region behind the incident shock are calculated from the Rankine-Hugoniot conditions for MHD shocks.  

Only four non-dimensional parameters characterize the initial configuration depicted by Figure~\ref{fig1}(a). 
The sonic Mach number $M$ parameterizes the incident shock velocity.
The contact discontinuity is expressed by the density jump $\rho_2/\rho_1$ and the ratio of the corrugation amplitude to the wavelength $\psi_0/\lambda$.  
The initial field strength is given by the plasma beta $\beta_0 = 8 \pi P_0/ B_0^2$, which is the ratio between the gas and magnetic pressures defined at the pre-shocked region.
Various situations can be examined only by choosing these four parameters; $M$, $\rho_2/\rho_1$, $\psi_0/\lambda$, and $\beta_0$.

\subsection{Basic Equations and Numerical Scheme}

The basic equations of ideal MHD are solved to study the interaction between the RMI and a magnetic field;
\begin{equation}
\frac{\partial \rho}{\partial t} + {\mbox{\boldmath{$\nabla$}}} \cdot
\left( \rho {\mbox{\boldmath{$v$}}} \right) = 0 \;,
\end{equation}
\begin{equation}
\frac{\partial \rho {\mbox{\boldmath{$v$}}}}{\partial t} +
{\mbox{\boldmath{$\nabla$}}} \cdot
\left[ \left( P + \frac{B^2}{8 \pi} \right) {\mbox{\boldmath{$I$}}}
+ \rho {\mbox{\boldmath{$v$}}}
{\mbox{\boldmath{$v$}}} - \frac{
{\mbox{\boldmath{$B$}}}{\mbox{\boldmath{$B$}}}}{4 \pi} \right] = 0 \;,
\end{equation}
\begin{equation}
\frac{\partial e}{\partial t} + {\mbox{\boldmath{$\nabla$}}} \cdot
\left[ \left( e + P + \frac{B^2}{8 \pi} \right) 
{\mbox{\boldmath{$v$}}} - \frac{
\left( {\mbox{\boldmath{$B$}}} \cdot {\mbox{\boldmath{$v$}}}
\right) {\mbox{\boldmath{$B$}}}}{4 \pi} 
\right] = 0 \;,
\end{equation}
\begin{equation}
\frac{\partial {\mbox{\boldmath{$B$}}}}{\partial t} = 
{\mbox{\boldmath{$\nabla$}}} \times
\left( {\mbox{\boldmath{$v$}}} \times {\mbox{\boldmath{$B$}}}
\right) \;,
\end{equation}
where $\rho$, $\bm{v}$, and $\bm{B}$ are the mass density, velocity, and magnetic field, respectively, and $e$ is the total energy density, 
\begin{equation}
e = \frac{P}{\gamma -1} + \frac{\rho v^2}2 + \frac{B^2}{8 \pi} \;.
\end{equation}
In our simulations, the equation of state for the ideal gas is adopted with the adiabatic index $\gamma = 5/3$. 

We solve the MHD equations by a conservative Godunov-type scheme \citep[e.g.,][]{sano98}.   
The hydrodynamical part of the equations is solved by a second-order Godunov method, using the exact solutions of a simplified MHD Riemann problem.    
The one-dimensional Riemann solver is simplified by including only the tangential component of a magnetic field.  
The characteristic velocity is the magnetosonic wave alone, and then the MHD Riemann problem can be solved in a way similar to the hydrodynamical one \citep{colella84}. 
The remaining terms, the induction equation and the magnetic tension part in the equation of motion, are solved by the consistent MoC-CT method \citep{stone92b,clarke96}, guaranteeing $\nabla \cdot \mbox{\boldmath $B$} = 0$ within round-off error throughout the calculation \citep{evans88}. 
The numerical scheme includes an additional numerical diffusion in the direction tangential to the shock surface to care for the carbuncle instability \citep{hanawa08}.  

Calculations are carried out in a frame moving with the velocity $v^{\ast}$ which is the interface velocity after the interaction with the incident shock.
For convenience, we define the locus of $y = 0$ as where the incident shock reaches the contact discontinuity at $t = 0$.
Then the post-shocked interface stays at $y = 0$ if it is not corrugated initially or stable for the RMI.
The simulations start before the incident shock hits the corrugated interface ($|Y_{\rm is} - Y_{\rm cd}| \gg \psi_0$).

The system of equations is normalized by the initial density and sound speed of the heavier fluid ``1'', $\rho_1 = 1$ and $c_{s1} = 1$, and the wavelength of the density fluctuation $\lambda = 1$.
The sound crossing time is also unity in this unit, $t_{s1} = \lambda / c_{s1} = 1$.
Most of the calculations in this paper use a standard resolution of $\Delta_x$ = $\Delta_y$ = $\lambda / 256$ unless otherwise stated.
A periodic boundary condition is used in the $x$-direction, and an outflow boundary condition is adopted in the $y$-direction. 
The size of the computational box in the $x$-direction is always set to be $L_x = \lambda$.
The $y$-length, on the other hand, is taken to be sufficiently wider, $L_y \geq 10 \lambda$, for the transmitted shock not to reach the edge of the computational domain. 

\section{Growth Velocity of RMI}

Figure~\ref{fig1}(b) is a schematic picture around the contact discontinuity after the shock passage. 
Now, the contact discontinuity moves with the velocity $v^{\ast}$. 
The pressure at the interface, $P^{\ast}$, becomes higher compared to the initial $P_0$. 
The incident shock is transmitted into the lighter fluid ``2'' ($y < Y_{\rm cd}$) which moves with the velocity $U_t (< 0)$, and the density behind the transmitted shock is $\rho^{\ast}_2$. 
Similarly, a reflected rarefaction forms with the sound speed at each position and expands the heavier fluid ``1'' to the density $\rho_1^{\ast}$ at the interface. 
If the interface is corrugated, the front surfaces of these two waves are also rippled.  
Then, refraction of the fluids at the wavefronts occurs as indicated by the arrows in Figure~\ref{fig1}(b).

Consider fluid motions near the shock surface in a frame moving with the transmitted shock or reflected rarefaction wave. 
For both waves, the upstream fluid moves along the $y$-axis.
However, the downstream motion should have the tangential component due to the obliqueness of the wavefront. 
The tangential shear motion at the contact discontinuity generates non-uniformity of the pressure perturbation across the interface, which will be the driving source of unstable growth for the RMI.  

The accurate formula for the asymptotic growth velocity of the RMI is derived by \citet{wouchuk96,wouchuk97}.   
In the Wouchuk-Nishihara (WN) model, the tangential velocities generated at the interface, $\delta v_1^{\ast}$ and $\delta v_2^{\ast}$ [see Figure~\ref{fig1}(b)], are the essential quantities for the linear growth of the RMI.   
Assuming a sinusoidal modulation of the interface, the tangential velocities at both sides of the interface can be calculated from the initial parameters.
Then, the linear growth velocity of the WN model is written as 
\begin{equation}
  v_{\rm lin} =
  \frac{\rho_{1}^{\ast} \delta v_{1}^{\ast}
      - \rho_{2}^{\ast} \delta v_{2}^{\ast}}
       {\rho_{1}^{\ast} + \rho_{2}^{\ast}} -
  \frac{\rho_{1}^{\ast} F_1 - \rho_{2}^{\ast} F_2 }
       {\rho_{1}^{\ast} + \rho_{2}^{\ast}} \;,
\label{eq:dvi}
\end{equation}
where the quantities $F_1$ and $F_2$ represent the sonic interaction between the contact surface and the transmitted and reflected waves, respectively, which are proportional to the amount of vorticity left behind the wavefronts in the bulk of each fluid.
For the case when a rarefaction is reflected, no vorticity is created in the expanded fluid, i.e., $F_1=0$. 

The growth velocity given by Equation~(\ref{eq:dvi}) is exact within the limits of linear theory and inviscid flow. 
It is valid for any initial configuration, and every element can be analytically calculated from the pre-shocked parameters \citep{wouchuk01a,wouchuk01b,coboscampos16,coboscampos17}.
The first term of the right-hand side of Equation~(\ref{eq:dvi}) is due to the instantaneous deposition of vorticity at the interface just after the shock interaction. 
The second term is non-negligible only for stronger shocks or highly compressible fluids and always reduces the growth velocity predicted by the first term. 
Within the linear regime ($\psi \ll \lambda$), the spike and bubble grow with the same velocity $v_{\rm lin}$.  
The negative growth velocity stands for the phase reversal that could occur in the rarefaction-reflected cases. 
Throughout our analysis, the WN formula $v_{\rm lin}$ is used as the linear growth velocity of the RMI for a given set of the parameters ($M$, $\rho_{2} / \rho_{1}$, $\psi_0 / \lambda$, and $\gamma$), since there is no corresponding theoretical linear model for MHD RMI.
Based on the impulse-driven linearized initial value problem, the pure hydrodynamic response dictates the initial interface behavior independent of the applied magnetic field \citep{shen20}.
Therefore, the choice of unmagnetized growth velocity would be rather appropriate for the definition of the Alfv{\'e}n number.

\subsection{Importance of Heavy-to-Light Configuration}

The RMI occurs for any Atwood number or any ratio of the interface densities.
When the adiabatic index $\gamma$ is assumed to be identical in both fluids, the light-to-heavy (heavy-to-light) configuration corresponds to the shock-reflected (rarefaction-reflected) case. 
In our previous work \citep{sano12}, the role of the magnetic field on the RMI has been examined intensely only for the shock-reflected cases.
However, the RMI can grow when the reflected wave is a rarefaction.
Thus we examine the rarefaction-reflected cases in this work and attempt to understand the magnetohydrodynamical evolutions of the RMI comprehensively.
In this subsection, we will demonstrate the importance of the rarefaction-reflected cases in a simple system.

\begin{figure*}
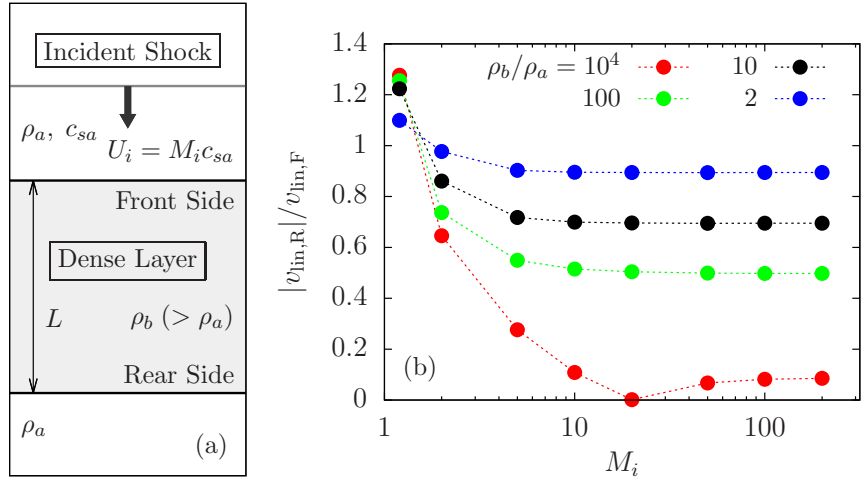

\begin{center}
\includegraphics[scale=0.9,clip]{fig2a.eps}%
\hspace{3mm}
{\includegraphics[scale=0.9,clip]{fig2b.eps}}%
\caption{
(a) 
A simple sketch of the shock passage through a dense layer.
Both sides of the layer are subject to the RMI.
(b)
The ratio of the growth velocity of the RMI between the front and rear sides shown as a function of the incident Mach number. 
The various density of the layer is considered from 2 to $10^4$.
The adiabatic index $\gamma = 5/3$ is assumed in this analysis.
\label{fig2}}
\end{center}
\end{figure*}

Let us suppose a situation when a planar shock front passes through a dense layer of the density $\rho_b$ ($> \rho_a$), which is depicted by Figure~\ref{fig2}(a).
Both sides of the dense layer are subjected to the RMI if it is corrugated.
The Mach number of the incident shock, $M_i = U_i / c_{sa}$, determines the growth velocity of the RMI at the front side, $v_{\rm lin, F}$, where $c_{sa}$ is the sound speed outside of the layer.
If the decay of the transmitted shock in the dense layer is negligible, the Mach number of the transmitted shock $M_t = U_t / c_{sb}$ characterizes the growth velocity at the rear surface, $v_{\rm lin,  R}$.
Assuming the same corrugation wavelength and amplitude at both sides, the ratio of the growth velocities can be derived as a function of the incident Mach number $M_i$ and the density ratio $\rho_b / \rho_a$.
Figure~\ref{fig2}(b) shows the ratio of the growth velocities at the front and rear sides of the dense layer for a given $M_i$.
For the cases of weaker shock $M_i \lesssim 2$, the growth velocity at the rear side exceeds that of the front side.
In general, when the density jump is small ($\rho_b / \rho_a \lesssim 10$), the RMI at the rear side is as fast as that at the front side, which suggests that both of the shock-reflected and rarefaction reflected RMI are important equally.
By contrast, the rear side instability becomes relatively unimportant when the density jump is high ($\rho_b / \rho_a \gtrsim 100$).
In the limit of high Mach number, larger density ratio brings slower growth velocity at the rear side.

\section{Numerical Results}

\subsection{Rarefaction-reflected cases}

The unstable motions of the RMI amplify ambient magnetic fields efficiently for the cases when a rarefaction wave is reflected.
This feature is almost identical to that seen in the RMI for the shock-reflected cases \citep{sano12}.

\begin{figure*}
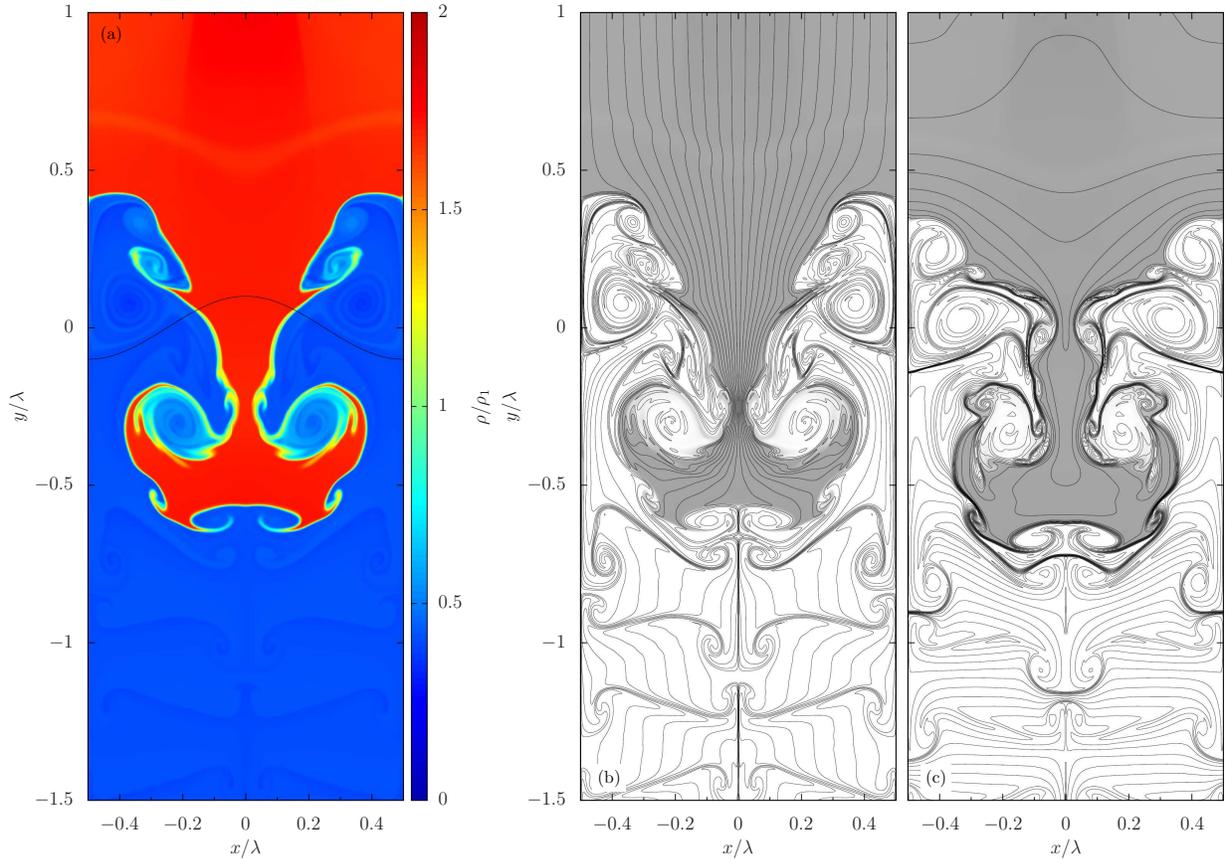

\begin{center}
\includegraphics[scale=0.6,clip]{fig3a.eps}%
\hspace{0.5mm}
{\includegraphics[scale=0.6,clip]{fig3b.eps}}%
\hspace{0.5mm}
{\includegraphics[scale=0.6,clip]{fig3c.eps}}%
\caption{
(a and b) 
Snapshot of (a) the density distribution and (b) the magnetic field lines in the late phase of the RMI growth taken at $k v_{\rm lin} t = 10$. 
The model parameters of this run is $M=100$, $\psi_0/\lambda = 0.1$, $\rho_1/\rho_2 = 0.1$, and $\beta_0 = 10^{8}$ with a uniform $B_y$ field.
The fluid evolution is solved in a frame moving with the interface velocity.
The initial location of the interface is shown by the solid black curve in (a).
(c) 
The spatial distribution of magnetic field lines in the perpendicular shock case started with an uniform $B_x$ field.
The initial parameters are the same as in (b) except for the magnetic field direction. 
The grid resolution used in these runs is $\Delta_x = \Delta_y = \lambda/512$.
\label{fig3}}
\end{center}
\end{figure*}

The interface is stretched with the magnetic field lines through the growth of the RMI.
Figures~\ref{fig3}(a) shows the density distribution at the nonlinear stage of the RMI, which is taken at $k v_{\rm lin} t = 10$. 
As for the initial conditions of this fiducial run, the density jump and corrugation amplitude of the interface are $\rho_2/\rho_1=0.1$ and $\psi_0/\lambda=0.1$, respectively. 
The adiabatic index is $\gamma = 5/3$ in both sides of the fluids.
For this case, the injected shock wave evolves into a transmitted shock and a reflected rarefaction after the interaction with the contact discontinuity.
The Mach number of the incident shock is assumed to be quite large, $M=100$.
The phase reversal of the surface modulation is a characteristic behavior for the rarefaction-reflected cases of the RMI. 
Then, the growth of the density spike in this run is towards the opposite direction to the initial pattern. 

A uniform magnetic field is applied in the $y$-direction, and then this is a parallel shock case with $\bm{B}_0 \parallel \bm{U}_i$. 
The initial plasma beta is set to be $\beta_0 = 10^8$.
The magnetic field develops passively to the RMI motions because it is not strong enough to alter the plasma flow. 
The perpendicular shock case ($\bm{B}_0 \perp \bm{U}_i$) with the same field strength is also performed, but the density structure formed by the RMI motions is quite similar except for tiny fluctuations associated with the mushroom-shaped spike. 

The magnetic field lines for the parallel and perpendicular shock cases are shown in Figures~\ref{fig3}(b) and \ref{fig3}(c).
The regions where the magnetic field lines are dense indicate the amplified magnetic field.
The strong magnetic fields appear near the interface on the side of the transmitted shock.
Small-amplitude fluctuations of the density appear in the lighter fluid, which is associated with reverse and cumulative jets left after the rippled transmitted shock \citep{stanic12,dell15}. 
These motions generate many filamentary structures of locally amplified magnetic fields between the interface and transmitted shock front. 
This feature can be recognized in both cases, suggesting it is independent of the field direction. 

\begin{figure*}
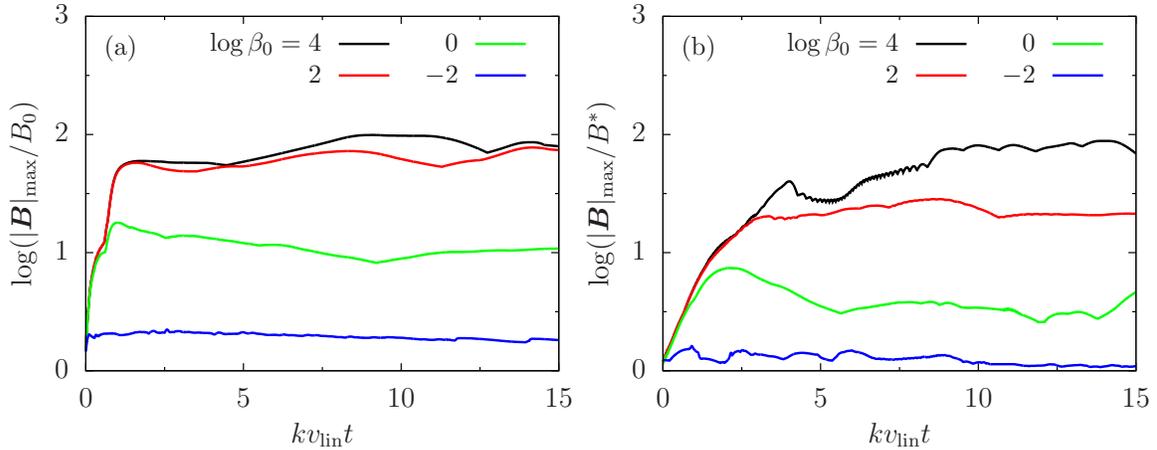

\begin{center}
{\includegraphics[scale=0.9,clip]{fig4a.eps}}%
\hspace{1mm}
{\includegraphics[scale=0.9,clip]{fig4b.eps}}%
\caption{
Time evolution of the maximum field strength $| \bm{B} |_{\max}$ shown as a function of the normalized time $k v_{\rm lin} t$ for (a) the parallel shock cases and (b) the perpendicular shock cases. 
The model parameters except for the initial magnetic field are the same as in the fiducial run depicted by Figure~\ref{fig3}.
\label{fig4}}
\end{center}
\end{figure*}

The magnetic field is amplified significantly in the timescale of the RMI, $t_{\rm rm}$.
Figures~\ref{fig4}(a) and \ref{fig4}(b) show the time evolution of the maximum strength of the magnetic field for the parallel-shock cases ($\bm{B}_0 \parallel \bm{U}_i$) and for the perpendicular-shock cases ($\bm{B}_0 \perp \bm{U}_i$), respectively.
The initial parameters except for the magnetic field strength are the same as in the fiducial run shown in Figure~\ref{fig3}.
The maximum strength is normalized by the initial strength of $B_0$ for the parallel-shock case, while it is divided by the post-shocked field strength $B^{\ast}$ in the fluid ``2'', which stands for the maximum field strength when the interface is flat.
The characteristic behavior of the maximum field strength shows little dependence on the direction of the initial magnetic field. 
If the ambient magnetic field is weak ($\beta_0 \gtrsim 10^4$), the amplification factor reaches two orders of magnitude. 
However, as the initial field strength increases, the maximum field strength decreases gradually. 
When the initial beta is $\beta_0 = 0.01$, no amplification is observed for both cases.
These features on the magnetic field amplification are qualitatively similar to the shock-reflected cases \citep{sano12}.
The strong magnetic fields are localized predominantly in thin filamentary structures. 
Thus the average magnetic energy over the entire region is at most a few times larger than the initial value.

\begin{figure}
\begin{center}
\includegraphics[scale=1.0,clip]{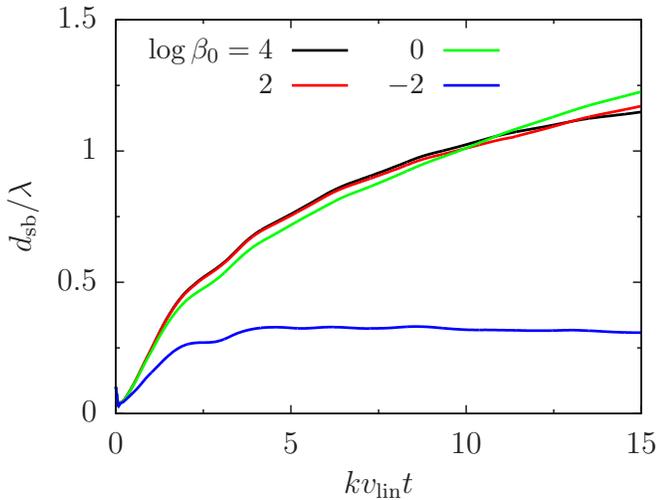}%
\caption{
The distance from the spike top to the bubble bottom $d_{sb}$ for the models with different initial plasma beta: $\beta_0 = 10^4$, $10^2$, 1, and 0.01.
The model parameters are identical to those of the parallel-shock runs shown in Figure~\ref{fig4}(a).
\label{fig5}}
\end{center}
\end{figure}

The magnetic field amplification is tightly linked to the growth of the RMI.
The interface is developed into a mushroom shape with smaller-scale vortical structures on the surface. 
The width of the mixing layer $d_{sb}$, which is defined by the distance from the spike top to the bubble bottom, increases with time. 
If the external magnetic field is large enough, then the growth of the RMI is suppressed significantly.
It is known that the critical field strength depends on the Mach number of the incident shock \citep{sano13}.
Figure~\ref{fig5} depicts the time history of the mixing width $d_{sb}$ to demonstrate the dependence on the external magnetic field. 
When the field strength is relatively weaker $\beta_0 \gtrsim 1$, the growth of the RMI is unaffected by the magnetic field.
However, if the plasma beta becomes $\beta_0 \sim 0.01$, the RMI is stabilized completely.
Then, the mixing layer shrinks into much smaller than that in the hydrodynamic case.
Although this figure includes only the parallel shock cases, the role of the magnetic field on the RMI suppression is independent of the initial field direction. 
Note that the critical plasma beta for the RMI is not given by unity. 

\subsection{Critical Alfv{\'e}n Number}

The critical value of the field strength varies depending on the model parameters, especially on the Mach number \citep{sano13, matsuoka17}. 
However, it is found that a single parameter characterizes magnetohydrodynamic evolutions of the RMI, that is, the Alfv{\'e}n number for the RMI.
The Alfv{\'e}n number is defined here as the ratio of the growth velocity to the Alfv{\'e}n speed,
\begin{equation}
R_{A} \equiv \frac{v_{\rm lin}}{v_A^{\ast}} \;,
\end{equation}
where $v_A^{\ast} = \min \{v_{A1}^{\ast},v^{\ast}_{A2}\}$ is the minimum of the Alfv{\'e}n speed at the interface after the shock passage.
The slower Alfv{\'e}n speed determines the criterion for the entire suppression of the fluctuations on both sides. 
It is turned out that the RMI growth is diminished when the Alfv{\'e}n number is less than unity, or when the Alfv{\'e}n speed exceeds the growth velocity in the unmagnetized limit.
The suppression condition $R_{A} \lesssim 1$ is almost the same meaning as that derived by \cite{sano13},
where the critical magnetic field is obtained by
\begin{equation}
\frac{B_{\rm crit}^2} {8 \pi P_0}
= \frac{\gamma}2 \alpha^2
\left( \frac{v_{\rm lin}}{c_{s}^{\ast}} \right)^2
\left( \frac{P^{\ast}}{P_0} \right) \;.
\end{equation}
This equation can be simplifed as $v_{A,{\rm crit}} = \alpha v_{\rm lin}$ where $\alpha$ is a factor of order unity, so that it is strictly related to the Alfv{\'e}n number.
Through a systematic survey of the broad parameter space in this paper, we find this condition is quite robust and valid for any parameters of the RMI, including the perpendicular shock cases and the parallel shock cases.

\begin{figure*}
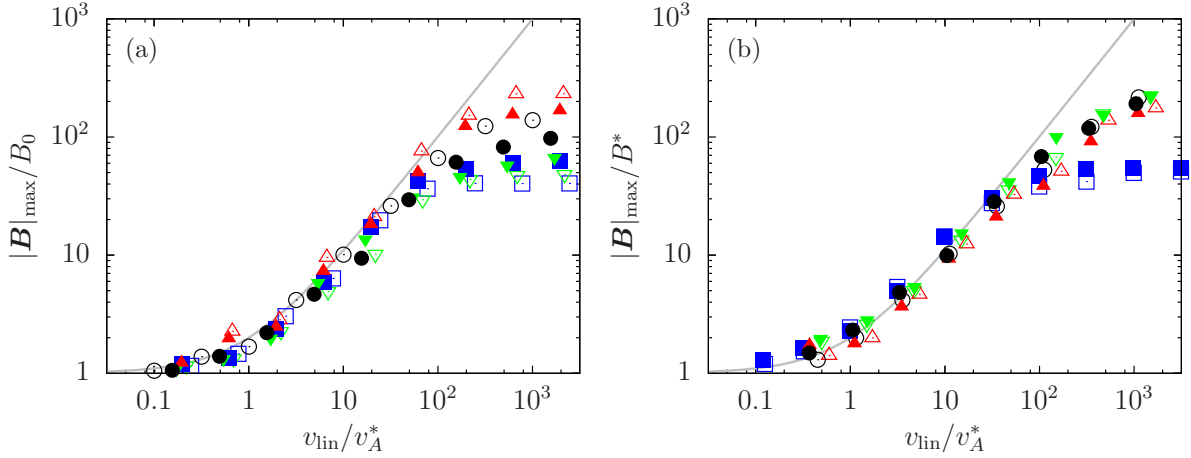

\begin{center}
\includegraphics[scale=0.9,clip]{fig6a.eps}%
\hspace{3mm}
{\includegraphics[scale=0.9,clip]{fig6b.eps}}%
\caption{
Dependence of the amplification factor of ambient magnetic field $|\mbox{\boldmath $B$}|_{\max} / B_0$ on the Alfv{\'e}n number $R_{A} = v_{\rm lin} / v_{A}^{\ast}$ for the cases when the magnetic field direction is (a) perpendicular and (b) parallel to the interface.
Filled (open) marks denote the density jump $\rho_2 / \rho_1$ is larger (smaller) than unity or when a shock (rarefaction) wave is reflected.
Black circles are models with $M=2$ and $\rho_2/\rho_1=10$ or $\rho_2/\rho_1=0.1$.  
Green (red) triangles are for $M=1.2$ ($M=10$) and $\rho_2/\rho_1=10$ or $\rho_2/\rho_1=0.1$ showing the dependence on the Mach number.
The RMI behavior on this diagram has little dependence on the density jump, which is demonstrated by the models of blue squares ($M=2$ and $\rho_2/\rho_1=2$ or $\rho_2/\rho_1=0.5$).
The initial perturbation amplitude in all the models shown in these figures is identical ($\psi_0 / \lambda_0 = 0.1$).
The gray curve indicates $|\mbox{\boldmath $B$}|_{\max} / B_0 = 1 + R_{A}$.
\label{fig6}}
\end{center}
\end{figure*}

Figure~\ref{fig6} shows the maximum magnetic field amplified by the RMI as a function of the Alfv{\'e}n number $R_A = v_{\rm lin} / v_A^{\ast}$ for various runs. 
The initial field direction is perpendicular to the interface for all the cases in Figure~\ref{fig6}(a).
The maximum field strength is measured at the end of calculations, where the magnetic field evolution is saturated sufficiently.
The amplification factor is normalized by $B_0$.
Even though the data include both the shock-reflected cases (filled marks) and rarefaction-reflected cases (open marks), the amplified magnetic field exhibits the same trend. 
When the Alfv{\'e}n number is less than unity, the amplification factor is almost unity.
No amplification denotes that the magnetic field stabilizes the RMI in this regime. 
However, as the Alfv{\'e}n number increases over unity, the maximum strength increases in proportion to $R_A$.
Then the maximum field energy of $B_{\max}$ becomes comparable to the turbulent kinetic energy of $v_{\rm lin}$ at the saturated state.
The amplification factor follows a relation of $1 + R_A$ at $R_A \lesssim 100$.
It should be emphasized that the $R_A$-dependence is valid for a wide range of parameters; The Mach number is from 1.2 to 10 and the density jump is from 0.1 to 10.

When $R_A \gtrsim 100$, the increase of the maximum strength is weakened obviously.
In this parameter range, the amplification factor is broadly scattered depending on the parameters, but it looks like a nearly flat function of $R_A$.
The amplified magnetic field seems to be higher as the incident Mach number increases.
However, the maximum field strength in this regime is found to depend on the numerical resolutions.
Then, it might be inappropriate to discuss quantitative features of the magnetic field only by our simulations performed in this paper.

The Alfv{\'e}n number is an excellent indicator of MHD RMI for the parallel shock cases too.
The amplification factors of the magnetic field are shown in Figure~\ref{fig6}(b) for various cases when the initial ambient field is in the $x$-direction.
The maximum field strength is made dimensionless by $B^{\ast}$ for this case.
The same trend as in Figure~\ref{fig6}(a) can be seen clearly in a range of $R_A \lesssim 100$.
To summarize, the RMI is suppressed by the tension force of the magnetic field when $R_A \lesssim 1$.
The surface modulations oscillate stably, and the mixing layer does not grow at all.
On the other hand, if the Alfv{\'e}n number exceeds unity, the magnetic field is enhanced through the interface stretching and tends to be toward the equipartition between $v_{A,\max} \approx v_{\rm lin}$.
The saturated level at $R_A \gtrsim 100$ indicates that the larger density jump causes the larger amplification in the parallel-shock runs.

\section{Discussions}

\subsection{Comparison with the other interfacial instabilities}

The linear growth of the RMI is reduced due to the presence of a strong magnetic field. 
The other similar instabilities, the Rayleigh-Taylor instability (RTI) and Kelvin-Helmholtz instability (KHI), are also stabilized by the magnetic field. 
Here we compare the dependence of the field strength on the critical wavelength to suppress these interfacial instabilities. 
We focus on the simplest situation where two uniform fluids are separated by a horizontal boundary and examine the effect of a uniform horizontal magnetic field $B$. 

If the gravity is working toward the lighter fluid ($\rho_2$) from the heavier one ($\rho_1$), it becomes unstable to the RTI. 
The linear growth rate of the RTI including the horizontal field is derived as 
\begin{equation}
\sigma_{\rm rt} = \left[ \frac{\rho_1 - \rho_2}{\rho_1 + \rho_2} g k 
- \frac{B^2}{2 \pi (\rho_1 + \rho_2)} k^2 \right]^{1/2} \;,
\end{equation}
\citep{chandrasekhar61} where $g$ is the gravitational acceleration and $k$ is the horizontal wave number.
The first and second terms on the right side are the growth rate for the unmagnetized case and the suppression rate originated from the Lorentz force. 
The RTI is stabilized by the magnetic field when
\begin{equation}
k > \frac{A g}{\bar{v}_A^2} \;,
\label{rti}
\end{equation}
where $A = (\rho_1 - \rho_2)/(\rho_1 + \rho_2)$ is the Atwood number, $\bar{v}_A = B / (4 \pi \bar{\rho})^{1/2}$ is the Alfv{\'e}n speed, and $\bar{\rho} = (\rho_1 + \rho_2)/2$ is the average density. 
Thus, the fluctuations in shorter wavelengths are more easily stabilized in terms of the RTI.

The interface is subject to the KHI, if the two fluids have relative horizontal motions ($v_1 \ne v_2$).
The linear growth rate is derived analytically as
\begin{equation}
\sigma_{\rm kh} = \left[ \frac{\rho_1 \rho_2}{\rho_1 + \rho_2} 
(v_1 - v_2)^2 k^2 - \frac{B^2}{2 \pi (\rho_1 + \rho_2)} k^2 
\right]^{1/2} 
\end{equation}
\citep{chandrasekhar61}.
Here a uniform magnetic field parallel to the interface is assumed.
The stabilization by the magnetic field is determined from the balance between the two terms on the right side of $\sigma_{\rm kh}$. 
Unlike the case of the RTI, the criterion is independent of the wavelength. 
The critical field strength is calculated as 
\begin{equation}
B_{c, {\rm kh}} = \sqrt{2 \pi \rho_1 \rho_2} | v_1 - v_2 | \;, 
\label{khi}
\end{equation}
for all wavelengths of perturbations.

The suppression condition for the RMI is obtained empirically by the numerical simulations. 
The growth of the RMI is dramatically reduced when the Alfv{\'e}n number is less than unity.
The linear growth velocity can be expressed by $v_{\rm lin} = \zeta k \psi_0 U_i$ using the incident shock velocity $U_i$.
The factor $\zeta$ is calculated from the pre-shocked quantities and takes roughly of the order of 0.1; $\zeta = \zeta (M, A, \gamma) \sim O(0.1)$.
Then, the RMI is suppressed by the magnetic field, when $v_{\rm lin} < v_{A}^{\ast}$.
This condition is rewritten as 
\begin{equation}
k < \frac{v_{A}^{\ast}}{\zeta \psi_0 U_i} \;.
\label{rmi}
\end{equation}
Therefore, for the RMI, the surface fluctuations with longer wavelengths are preferentially stabilized as oppose to the RTI. 

The characteristics of the stabilized wavelength are distinctly different for each instability.
To be precise, though, this condition should depend on the direction of the magnetic field.
Especially, interchange modes are unaffected at all by the restoring force of magnetic fields.
However, the relations between the critical wavelength and magnetic field [Equations~(\ref{rti}), (\ref{khi}), and (\ref{rmi})] are beneficial to intuitively understand the role of the magnetic field for each instability.

 \subsection{Nonlinear evolution of multi-mode RMI}

Single-mode analysis has the advantage of understanding the nonlinear features of the RMI in a clear picture. 
However, the actual situations are not always as simple as such a system.
The growth process of more complex multi-mode disturbances is briefly touched upon in the last section.

The interface modulation for our multi-mode analysis is assumed to be 
\begin{equation}
y = Y_{\rm cd} + \frac{\psi_0}{\sqrt2 \sigma}
\sum_{n=1}^{N} a_n \cos \left[ 2 \pi \left( \frac{x}{\lambda_n} +
  \phi_n
\right) \right] \;,
\label{ycd}
\end{equation}
where $\lambda_n = {\lambda}/{n}$ is the multi-mode wavelength, $a_n$ and $\phi_n$ are random coefficients between 0 and 1, $\sigma$ is the standard deviation of the summation part to normalize the fluctuation to be $1 / \sqrt2$, and the initial amplitude $\psi_0$ determines the size of the perturbation.
The single-mode analysis corresponds to the case of $N=1$.
The multi-mode corrugation with $N=10$, for example, consists of ten different wavelengths of $n = 1$--10.

\begin{figure*}
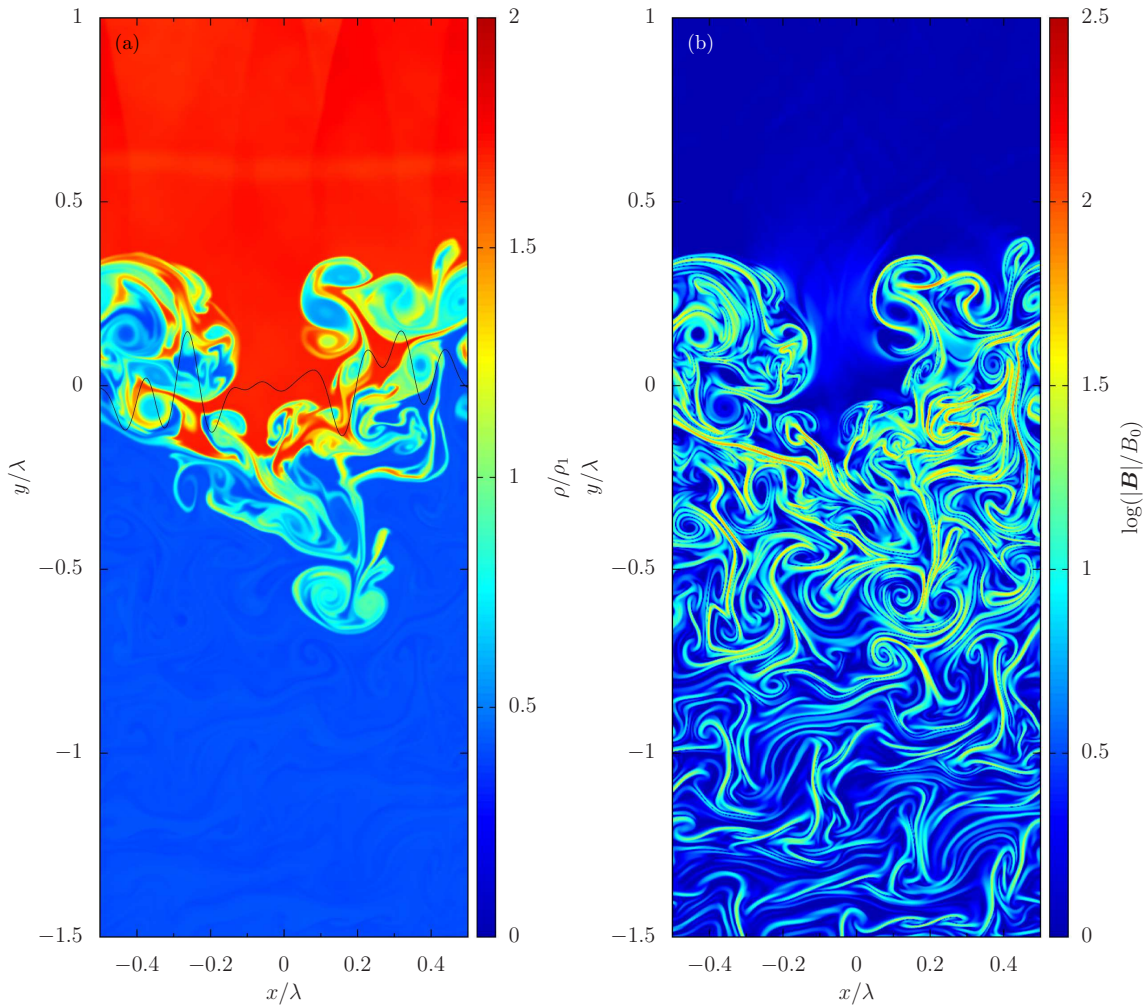

\begin{center}
\includegraphics[scale=0.7,clip]{fig7a.eps}%
\hspace{0.5mm}
{\includegraphics[scale=0.7,clip]{fig7b.eps}}%
\caption{
Snapshot of (a) the density distribution and (b) the magnetic field strength in the late phase of the multi-mode RMI growth taken at $k v_{\rm lin} t = 10$. 
The model parameters of this run is $M=100$, $\rho_1/\rho_2 = 0.1$, and $\beta_0 = 10^{8}$ with a uniform $B_y$ field.
The multi-mode surface modulation is given by Equation~(\ref{ycd}) with $\psi_0/\lambda = 0.1$ and $N = 10$.
The initial location of the interface is shown by the solid black curve in (a).
The grid resolution used in this run is $\Delta_x = \Delta_y = \lambda/512$.
\label{fig7}}
\end{center}
\end{figure*}

The magnetic field amplification for the case of multi-mode fluctuations is examined in Figure~\ref{fig7}.
The model parameters are mostly the same as the fiducial run of the single-mode case; $M = 100$ and $\rho_2 / \rho_1 = 0.1$.
The position of the  initial interface is given by Equation~(\ref{ycd}) with $\psi_0 / \lambda = 0.1$ and $N = 10$.
As a result of the RMI growth, the interface shows a complicated shape where various-scale mushroom-structures are growing randomly in all directions. 
However, the thickness of the mixing layer is comparable to that in the corresponding single-mode case shown by Figure~\ref{fig3}.

\begin{table*}
\caption{
Parameters for the initial surface modulation in multi-mode analysis.
\label{tab1}}
\startlongtable
\begin{tabular}{cccccccccccc}
\hline \hline
$N$ & $\sigma$ & $\phi_1$ & $\phi_2$ & $\phi_3$ & $\phi_4$ & $\phi_5$ &
$\phi_6$ & $\phi_7$ & $\phi_8$ & $\phi_9$ & $\phi_{10}$ \\ \hline
10 & 2.2360 & 0.0919 & 0.5297 & 0.3835 & 0.0668  & 0.6711 & 0.6316 
& 0.5194 & 0.2377 & 0.7621 & 0.9092 \\
3 & 1.2247 & 0.0919 & 0.5297 & 0.3835 & - & - & - & - & - & - & - \\
1 & 0.7071 & 1.0000 & - & - & - & - & - & - & - & - & - \\
\hline \hline
\end{tabular}
\end{table*}

\begin{figure}
\begin{center}
\includegraphics[scale=1.0,clip]{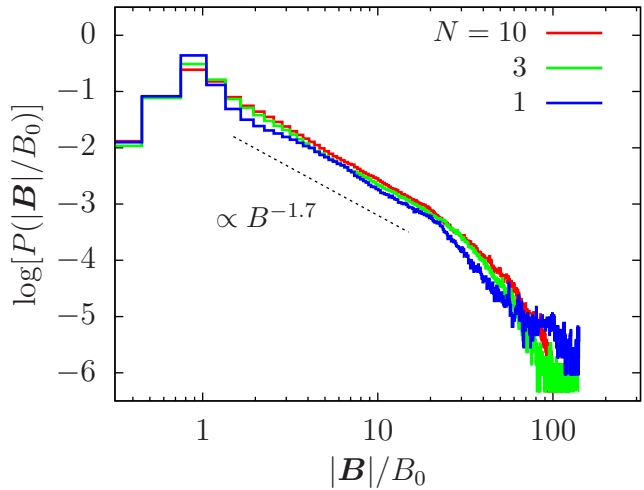}%
\caption{
Probability distribution functions of the magnetic field strength at the nonlinear regime of the multi-mode RMI ($N = 10$ and 3) and the single-mode case ($N = 1$). 
The model parameters are the same as in Figure~\ref{fig7} except for the shape of the surface fluctuations. 
\label{fig8}}
\end{center}
\end{figure}

The initial magnetic field is relatively weak as $\beta_0 = 10^8$ so that the RMI motions enhance the seed field in this run. 
The amplified field structure exhibits that a large number of curled filaments fill the entire area between the interface and the transmitted shock front [see Figure~\ref{fig7}(b)]. 
The filamentary feature is caused by the complicated vortex structure left behind the transmitted shock. 
Figure~\ref{fig8} indicates the probability distribution function of the magnetic field strength for the cases of $N = 10$, 3, and 1.
The parameters for the interface modulation are listed in Table~\ref{tab1}.
Here we assume $a_n = 1$ for a white noise amplitude.
The maximum field strength in the multi-mode runs is nearly the same as that in the single-mode. 
Furthermore, the power-law feature seen in the amplified component is shared by the single-mode and multi-mode runs. 
The index fitted in a range of $2 < | \bm{B} | / B_0 < 20$ is about $-1.7$, which may be the characteristic quantity of the RMI turbulence driven by a shock wave.
The power-law features are unaffected by choice of the random numbers for the initial interface shape in Equation (\ref{ycd}).
It should be noticed that a similar power index is obtained in interstellar turbulence simulations by \cite{inoue09}, indicating the RMI motions could contribute to the evolutions of interstellar magnetic fields. 
However, three-dimensional simulations will be essential for further quantitative comparisons with observations.

\section{Summary}

We have investigated the role of a magnetic field in the nonlinear evolution of the RMI using two-dimensional MHD simulations.
An essential feature of the RMI is that the shock wave can excite the unstable motions when it propagates from the low-density side of the contact discontinuity to the high-density side and vice versa. 
In this paper, we intensively analyze the MHD RMI in the heavy-to-light configuration.
It is found that the characteristic features of MHD RMI are well described by the Alfv{\'e}n number for the RMI that is the ratio of the linear growth velocity to the Alfv{\'e}n speed.
The critical condition on the Alfv{\'e}n number is $R_A \sim 1$, which distinguishes the suppression of the RMI and the field amplification by the RMI.
The obtained criterion is applicable universally for the heavy-to-light and light-to-heavy configurations.
It is also independent of any initial parameters such as the incident Mach number, density jump, and the magnetic field geometry. 

The magnetic features amplified by the RMI growth indicate similarities to those seen in the simulations of interstellar turbulence.
This fact suggests that the RMI-like behaviors excited by supernova-shock propagation through inhomogeneous density distribution may contribute to the evolution and amplification of the interstellar magnetic fields.
Therefore, further deep comparison between the RMI and interstellar turbulence would be a curious subject to be studied in the future.

\begin{acknowledgments}
We thank C. Matsuoka, K. Nishihara, and K. Shibata for useful discussions and encouragement.
This work was partly supported by JSPS KAKENHI Grant Number JP26287147, HPCI Systems Research Projects (Project ID hp120227), and joint research project of ILE, Osaka University.
Computations were carried out on SX-8R, SX-ACE at the Cybermedia Center, and SX-9B, SX-ACE at the Institute of Laser Engineering of Osaka University. 
\end{acknowledgments}

\end{document}